\documentclass[12pt]{article}
\usepackage{amssymb,slashed,latexsym,amsmath}
\usepackage[normalem]{ulem}
\newcommand{\ft}[2]{{\textstyle\frac{#1}{#2}}}

\def\rmi{{\rm i}}
\def\rmd{{\rm d}}
\newsavebox{\uuunit}
\sbox{\uuunit}
    {\setlength{\unitlength}{0.825em}
     \begin{picture}(0.6,0.7)
        \thinlines
        \put(0,0){\line(1,0){0.5}}
        \put(0.15,0){\line(0,1){0.7}}
        \put(0.35,0){\line(0,1){0.8}}
       \multiput(0.3,0.8)(-0.04,-0.02){12}{\rule{0.5pt}{0.5pt}}
     \end {picture}}
\newcommand {\unity}{\mathord{\!\usebox{\uuunit}}}

\csname @addtoreset\endcsname{equation}{section}
   \usepackage[nosort]{cite}
  \usepackage[pdftex]{hyperref}
\usepackage{tcolorbox}

\newcommand{\dr}{\raise.3ex\hbox{$\stackrel{\leftarrow}{\delta  }$}{}}
\newcommand{\dl}{\raise.3ex\hbox{$\stackrel{\rightarrow}{\delta }$}{} }
\newcommand{\pl}{\raise.3ex\hbox{$\stackrel{\rightarrow}{\partial }$}{} }
\newcommand{\ualpha}{{\underline{\alpha}}}
\newcommand{\ubeta}{{\underline{\beta}}}
\newcommand{\ugamma}{{\underline{\gamma}}}
\newcommand{\umu}{{\underline{\mu}}}
\newcommand{\unu}{{\underline{\nu}}}
\newcommand{\uX}{{\underline{X}}}
\newcommand{\uF}{{\underline{F}}}

\begin{document}

\begin{titlepage}
\phantom{.}
\vspace{.5cm}
\begin{center}
\baselineskip=16pt
{\LARGE    Covariant field equations in supergravity 
}\\
\vfill
{\large Bram Vanhecke $^{1,2}$ and Antoine Van Proeyen $^1$  
  } \\
\vfill
{\small $^1$ KU Leuven, Institute for Theoretical Physics,\\
       Celestijnenlaan 200D B-3001 Leuven, Belgium.\\[5mm]
       $^2$ Ghent University, Faculty of Physics,\\ Krijgslaan 281, 9000 Gent, Belgium
      \\[2mm] }
\end{center}
\vfill
\begin{center}
{\bf Abstract}
\end{center}
{\small Covariance is a useful property for handling supergravity theories. In this paper, we prove a covariance property of supergravity field equations:
 under reasonable conditions, field equations of supergravity are covariant modulo other field equations.
 We prove that for any supergravity there exist such covariant equations of motion, other than the regular equations of motion, that are equivalent to the latter. The relations that we find between field equations and their covariant form can be used to obtain multiplets of field equations.
 In practice, the covariant field equations are easily found by simply covariantizing the ordinary field equations.
 }\\[1cm]
 keywords: field theory; symmetries; supergravity
 \vspace{2mm} \vfill \hrule width 3.cm
{\footnotesize \noindent e-mails:
bavhecke.vanhecke@ugent.be; antoine.vanproeyen@fys.kuleuven.be }
\end{titlepage}
\addtocounter{page}{1}
 \tableofcontents{}
\newpage
\section{Introduction}
A typical supergravity action can easily fill a few lines to an entire page, depending on the density of notation. From the beginning it was therefore clear there was a need for new mathematical formalisms that might provide general proofs and properties to simplify calculations and provide general insight into supergravity. A useful property to simplify equations is covariance. This property is already quite central in general relativity. The only quantities that can hold observer-independent meaning are covariant quantities, or tensors, as they are called in the general relativity framework. The definition of a covariant quantity was generalised in supergravity to a more general concept. If some quantity is known to be covariant, its calculation can be tremendously simplified.

For quite some time it has been hypothesised that field equations in supergravity are (super-)covariant quantities, modulo other field equations.\footnote{`Field equations' or `equations of motion' (eom) are used interchangeably. They are the classical Euler-Lagrange equations derived from a Lagrangian.  When we discuss below `covariance of field equation' we intend the covariance of the expression $T_i$ in the field equation $T_i\approx 0$.}
This has already been used in \cite{Cremmer:1978km} to fix coefficients in the action. However, the straightforward definition of field equations does in general not lead to covariant equations. Therefore we have analysed the relation between Euler-Lagrange equations and covariant expressions in general. We will find out what this relation exactly means and why this is valid. We will prove that under conditions satisfied for a large class of supergravity theories, there is a set of covariant equations that have the same solutions as the field equations. Furthermore, this set of covariant equations of motion can easily be found by covariantizing the Euler-Lagrange equations.

The relations that we find between straightforward field equations and covariant field equations is useful in the calculation of their supersymmetry transformations and thus for determining their inclusion in supermultiplets \cite{Ferrara:2017yhz}. The multiplets of field equations and currents are e.g. used in the recently developed localization techniques in supersymmetric quantum field theories \cite{Festuccia:2011ws, Dumitrescu:2012ha, Pestun:2016zxk, Pufu:2016zxm}.

One may confront this research with the superspace approach \cite{Wess:1977fn}.\footnote{Useful books on superspace methods are \cite{Gates:1983nr,Buchbinder:1998qv,Binetruy:2000zx}.} For ${\cal N}=1$, $D=4$, in the superspace book \cite{Gates:1983nr},  Gates et al. discuss in Sec. 5.3 a `Covariant approach to supergravity'. After introducing convenient formulation in terms of covariant variations they arrive for this case after long considerations to covariant field equations in Sec.5.3.j.
It is expected but not proven that such a procedure is always possible.
An obstacle is the fact that many formulations in superspace are based on on-shell superfields. Therefore, equations of motion should be already imposed before deriving Euler-Lagrange equations from an action.\footnote{One noteworthy case is IIB supergravity in 10 dimensions \cite{Schwarz:1983wa,Schwarz:1983qr,Howe:1984sr}. The consistent superspace geometry where only conventional constraints were applied, leads then directly to on-shell supergravity. In this case, also component formulations starting only from Euler-Lagrange equations of an action are difficult. One way out is the action based on the Pasti-Sorokin-Tonin mechanism  \cite{Pasti:1997vs}, which involves intrinsic nonlinearities and extra gauge symmetries. }
One probably needs to define various superspaces: e.g. superspaces extended by extra coordinates for hypermultiplets \cite{Galperin:1984av,Lindstrom:1987ks,Lindstrom:1989ne,Galperin:2001uw,Kuzenko:2008ep} or based on pure spinors for $D=11$  \cite{Cederwall:2009ez,Cederwall:2010tn}. In most of the superspace formulations constraints determine covariant equations, but they do not follow from an action principle.
Although powerful in principle, it is therefore not obvious how a general proof of covariant Euler-Lagrange equations in superspace, valid for all supergravities, can be developed.

The aim of this paper is to develop such a proof based on component formulations. The action contains non-covariant quantities, but we prove that the equations of motions can always be made covariant.

We will start in section \ref{ss:examples} with examples in ${\cal N}=1$, $D=4$ and in $D=11$ supergravity. They illustrate that a lot of technical manipulations are necessary to get from the action to a covariant form of field equations. Furthermore, we see that for obtaining the covariant field equation of the Maxwell field one needs to use the field equation of the gaugino, which we will later explain from the general proof of our theorems.
In section \ref{ss:basic} we will first mention a general result on the symmetry transformations of field equations. We discuss this first for general coordinate transformations and then the other transformations, indicated as `standard gauge transformations'. In section \ref{ss:coveom} we discuss the structure of covariant eom, first for a typical case with just two fields.
This is followed in section \ref{ss:coveomgen} with the analysis of the general case leading to the main result of the paper, which is the statement under which conditions one can construct the covariant set of equations of motion. In section \ref{ss:ConvenientStructure} we explain a common structure of supergravity theories that allows to solve these conditions and explain how the covariant equations are obtained. To conclude we summarize the result and the conditions under which the theorems apply in section \ref{ss:conclusions}.

In appendix \ref{app:covariant} we summarize the rules and definitions of covariant quantities, as they have been explained in \cite{Freedman:2012zz}. Appendix \ref{app:exchiralg} specifies the abstract quantities, used in the general proofs, for the example of the chiral multiplet.

\section{Examples}
\label{ss:examples}
\subsection{Super-Maxwell-Einstein theory}
\label{ss:sME}
We consider here the coupling of the ${\cal N}=1$ super-Maxwell theory, with gauge field $A_\mu $ and gaugino $\lambda $, coupled to supergravity with frame field $e_\mu ^a$ and gravitino $\psi _\mu$. This theory has been constructed in \cite{Ferrara:1976um,Ferrara:1976ni,Freedman:1976ej}.
We will use the notation of \cite{Freedman:2012zz}. The Lagrangian is
\begin{align}
e^{-1}{\cal L}=&\frac1{2\kappa ^2}\left[ R(\hat{\omega })- \bar \psi_\mu\gamma ^{\mu \nu \rho }D_\nu \psi _\rho  \right]
 -\ft14 F^{\mu \nu} F_{\mu \nu } -\ft12 \bar \lambda  \slashed{D}^{(0)}\lambda+\ft18\bar \psi _\mu \gamma ^{ab}
\left( F_{ab}+ \widehat F_{ab} \right)
\gamma ^\mu \lambda \nonumber\\ &+ \ft{1}{32}\rmi e^{-1}\varepsilon^{\mu\nu\rho\sigma}\bar
 \psi _\mu \gamma_\nu \psi _\rho\bar \lambda \gamma_*\gamma _\sigma \lambda+ \ft{3}{64}\kappa^2 (\bar{\lambda}\gamma_a\gamma_*\lambda )(\bar{\lambda}\gamma^a\gamma_*\lambda )
\,.\label{phenomL}
\end{align}
We distinguish here between the spin connection with and without gravitino torsion
\begin{align}
  \hat{\omega } _\mu{}^{ab}=&\omega _\mu{}^{ab}(e) + K_\mu {}^{ab}\,,\nonumber\\
  &\omega _\mu {}^{ab}(e)=2 e^{\nu[a} \partial_{[\mu} e_{\nu]}{}^{b]} -
e^{\nu[a}e^{b]\sigma} e_{\mu c} \partial_\nu e_\sigma{}^c\,,\qquad
K_\mu {}^{ab}= \ft12\bar \psi _\mu \gamma ^{[a}\psi ^{b]}+\ft14\bar \psi ^a\gamma _\mu \psi ^b\,.
 \label{covomega}
\end{align}
This determines $R(\hat{\omega } )$, and
\begin{align}
D^{(0)}_\mu \lambda  =&\left( \partial _\mu +\ft14 \omega _\mu{}^{ab}(e)\gamma _{ab} \right)\lambda \,,\nonumber\\
D_\nu \psi _\rho=&\left( \partial _\nu +\ft14 \hat{\omega } _\nu
{}^{ab}\gamma _{ab} \right)\psi _\rho  \,,\nonumber\\
\widehat{F}_{ab}=&e_a{}^\mu e_b{}^\nu\left( F_{\mu \nu }+\bar \psi _{[\mu }  \gamma_{\nu ]}\lambda \right)\,,\qquad  F_{\mu \nu }=2\partial _{[\mu }A_{\nu ]}\,.
 \label{hatFsugra}
\end{align}
The action is invariant under the local supersymmetry and gauge transformation rules
\begin{align}
\delta e^a_\mu = &\ft12 \bar{\epsilon}\gamma^a\psi_\mu\,,\nonumber\\
\delta \psi_\mu =& \left(\partial_\mu +\ft14 \hat{\omega }_\mu{}^{ab}\gamma_{ab} + \Xi _\mu \right)\epsilon\,,\nonumber\\
\delta A_\mu =& -\ft12\bar{\epsilon}\gamma_\mu \lambda +\partial _\mu \theta \,,\nonumber\\
\delta \lambda =& \ft14 \gamma^{ab}\widehat{F}_{ab}\epsilon\,,
\label{gauge multiplet variations}
\end{align}
where $\Xi _\mu $ is the contribution due to the fact that the auxiliary vector field of supergravity is eliminated:
\begin{equation}
\Xi _\mu =\ft12\rmi\left(\gamma _\mu \gamma ^a -3e_\mu ^a\right)\gamma _* A_a^{\rm F}\,,\qquad A_a^{\rm F}=-\rmi\ft18\kappa ^2\bar \lambda \gamma _a\gamma _*\lambda \,.
\end{equation}
Observe that $\widehat{F}_{ab}$ is the covariant curvature following the rule in (\ref{covcurvature}), since the quantity ${\cal M}_{\mu B}{}^A $ (where $B$ is an unwritten spinor index and $A$ refers to the symmetry with parameter $\theta $) is by comparison of  (\ref{gauge multiplet variations}) and  (\ref{modifiedGaugeTrA})
\begin{equation}
  {\cal M}_\mu =  -\ft12\gamma_\mu \lambda\,.
 \label{MmuME}
\end{equation}
Following the rule  (\ref{covariant derivative}), a fully covariant derivative of $\lambda $ is
\begin{equation}
  {\cal D}_\mu \lambda =\left( \partial _\mu +\ft14 \hat{\omega } _\mu
{}^{ab}\gamma _{ab} \right)\lambda -\ft14\gamma ^{ab}\widehat{F}_{ab}\psi _\mu \,.
 \label{covderlambda}
\end{equation}

When we calculate the field equation of $\lambda $, we have to be careful also with the torsion terms and obtain
\begin{align}
e^{-1}\frac{\dl S}{\delta \bar{\lambda}}=&-\slashed{D}^{(0)}\lambda
+ \ft18\gamma^\mu\gamma^{ab}\Big(F_{ab}+\widehat{F}_{ab}\Big)\psi_\mu-\ft18 \gamma_{[b}\psi _{a] }\bar \psi _\mu \gamma ^{ab}\gamma ^\mu \lambda\nonumber\\
& +\ft{1}{16}\rmi \gamma_*\gamma_\sigma \lambda e^{-1} \varepsilon^{\mu\nu\rho\sigma}\bar{\psi}_\mu\gamma_\nu\psi_\rho +\ft{3}{16}\kappa ^2  \gamma _a\gamma _*\lambda \bar \lambda \gamma ^a\gamma _*\lambda\,.
\label{lambda eom hard}
\end{align}
One rewrites this in terms of the covariant object ${\cal D}_a\lambda $:
\begin{align}
  e^{-1}  \frac{\delta S}{\delta \bar \lambda }=&-\slashed{\cal D}\lambda
+\ft14\gamma ^\mu K_\mu {}^{ab}\gamma _{ab}\lambda -\ft18 \gamma ^\mu  \gamma ^{ab}
\left( \widehat F_{ab}- F_{ab} \right)\psi _\mu
-\ft18  \gamma_{[b}\psi _{a] }\bar \psi _c \gamma ^{ab}\gamma ^c \lambda \nonumber\\
&- \ft{1}{16} \gamma ^{abc} \lambda\bar
 \psi _a \gamma_b \psi _c +\ft3{16}\kappa ^2\gamma _a\gamma _*\lambda \bar \lambda \gamma ^a\gamma _*\lambda \nonumber\\
 =&-\slashed{\cal D}\lambda+\ft3{16}\kappa ^2\gamma _a\gamma _*\lambda \bar \lambda \gamma ^a\gamma _*\lambda\,.
 \label{felambdastep1}
\end{align}
One needs a Fierz transformation and some $\gamma $-matrix manipulations to show that all the non-covariant terms cancel.
If we would have known, we could have saved ourselves a lot of trouble ...

Let us then calculate the field equation of the gauge field. The straightforward calculation gives
\begin{align}
\frac{\delta S}{\delta A^\mu}&= \partial^\nu \bigg( eF_{\nu \mu}-e\ft{1}{2}\bar{\psi}_\sigma\gamma_{\nu\mu}\gamma^\sigma\lambda \bigg)\nonumber\\
&=ee_\mu^aD^{(0)b} \bigg( F_{ba}-\ft{1}{2}\bar{\psi}_\sigma\gamma_{ba}\gamma^\sigma\lambda \bigg)= ee_\mu^aD^{(0)b}\bigg(\widehat{F}_{ba}-\ft{1}{2}\bar{\psi}_c{\gamma_{ba}}^c\lambda\bigg)\,.
\label{feAmu}
\end{align}
In order to know what is ${\cal D}^b\widehat{F}_{ba}$, which we expect in a covariant expression, we have to know first the supersymmetry transformation of $\widehat{F}_{ba}$. Since we know that the latter should be a covariant quantity, we can easily obtain from (\ref{gauge multiplet variations}):
\begin{equation}
  \delta  \widehat{F}_{ab} =\bar{\epsilon}\gamma_{[a}\mathcal{D}_{b]}\lambda+\bar{\lambda}\gamma_{[a}\Xi_{b]}\epsilon\,.
 \label{delhatF}
\end{equation}
This implies that the covariant ${\cal D}^b\widehat{F}_{ab}$ is of the form
\begin{equation}
  \mathcal{D}^b \widehat{F}_{ba} =  D^b \widehat{F}_{ba} -\bar{\psi}^b\gamma_{[b}\mathcal{D}_{a]}\lambda-\bar{\lambda}\gamma_{[b}\Xi_{a]}\psi^b \,.
\end{equation}
These derivatives have torsion, i.e.
\begin{equation}
  D^b \widehat{F}_{ba}= D^{(0)b}\widehat{F}_{ba}- 2 K^b{}_{[a}{}^c\widehat{F}_{b]c}\,.
 \label{DhatF}
\end{equation}
Furthermore, we see that (\ref{feAmu}) contains the derivative of the gravitino, which should be part of a covariant curvature
\begin{equation}
  \widehat{R(Q)}_{ab}= e_a {}^\mu  e_b{}^\nu \widehat{R(Q)}_{\mu \nu}\,,\qquad \widehat{R(Q)}_{\mu \nu}=2 D_{[\mu}\psi_{\nu]}+2 \Xi _{[\mu}\psi _{\nu]}\,.
 \label{covRQ}
\end{equation}
Taking all this into account we find after a lot of manipulations
\begin{equation}
  e^{-1}e^\mu_a\frac{\delta S}{\delta A^\mu}= \mathcal{D}^b \widehat{F}_{ba} -\ft{1}{4}\bar{\lambda}{\gamma_{bac}} \widehat{R(Q)}^{bc} +\ft{1}{2}\bar{\psi}_\mu \gamma^\mu {}_{a}\left(\slashed{\mathcal{D}}\lambda-\ft3{16}\kappa ^2\gamma ^d\gamma _*\lambda\bar \lambda \gamma _d\gamma _*\lambda  \right)\,.
  \label{feAmucov}
\end{equation}
It is clear that the first two terms are covariant while the last term is not, but is proportional to the eom of $\lambda$. Hence we find that not all field equations are covariant, but the non-covariant part is proportional to a field equation. So we could have written that the field equations are equivalent to the two covariant equations
\begin{align}
  & -\slashed{\cal D}\lambda  +\ft3{16}\kappa ^2\gamma _a\gamma _*\lambda \bar \lambda \gamma ^a\gamma _*\lambda \approx 0\,,\nonumber\\
    & \mathcal{D}^b \widehat{F}_{ba} -\frac{1}{4}\bar{\lambda}{\gamma_{bac}} \widehat{R(Q)}^{bc} \approx 0\,.
\label{covfeSME}
\end{align}

\subsection{Including the auxiliary field}

The previous model with the auxiliary field involves a few simple modifications. The action would have an extra term $+\ft14D^2$ where $D$ is the auxiliary field, which thus has the (covariant) field equation
\begin{equation}
  D\approx 0\,.
 \label{Dapprox0}
\end{equation}
However, since the transformation  rules change:
\begin{equation}
  \delta \lambda = \ldots  + \ft12\rmi\gamma _* D\epsilon \,,\qquad \delta D = \ft12\rmi\bar \epsilon \gamma _*\slashed{\cal D}\lambda \,,
 \label{delextraD}
\end{equation}
the covariant derivative of $\lambda $ has an extra term
\begin{equation}
  {\cal D}_\mu \lambda =\ldots  -\ft12\rmi\gamma _*\psi _\mu D\,.
 \label{DmulambdaD}
\end{equation}
Therefore, if we write the field equation of the gaugino in the form (\ref{felambdastep1}), we have to compensate it with the term in (\ref{DmulambdaD}). Thus, with the full covariant derivative, we write now
\begin{equation}
    -e^{-1}  \frac{\delta S}{\delta \bar \lambda }=\slashed{\cal D}\lambda
+ \ft{1}{16} \gamma ^{abc} \lambda\bar
 \psi _a \gamma_b \psi _c -\ft3{16}\kappa ^2\gamma _a\gamma _*\lambda \bar \lambda \gamma ^a\gamma _*\lambda +\ft12\rmi\gamma ^\mu \gamma _*\psi _\mu D\,.
 \label{felambdaD}
\end{equation}
The last term is not covariant, but is proportional to  (\ref{Dapprox0}). The same modification should be done to (\ref{feAmucov}). However, the covariant equations are still (\ref{covfeSME}), supplemented with (\ref{Dapprox0}).

\subsection{The chiral multiplet in supergravity}
\label{ss:exchiral}
The coupling of the chiral multiplet to supergravity \cite{Ferrara:1976ni,Ferrara:1976kg,Das:1977pu} is another useful example. In the notation of
\cite{Freedman:2012zz}, we consider the superconformal-invariant action of a chiral multiplet $\{X,\,\Omega ,\,F\}$ coupled to the Weyl multiplet\footnote{We already omit here $b_\mu $, the gauge field of dilatations, as the standard gauge choice of special conformal transformations puts it to zero. It could be included, but would not change anything to the discussion.} $\{e_\mu ^a,\,\psi _\mu ,\, A_\mu \}$ with a minimal kinetic term and arbitrary superpotential:
\begin{equation}
  S= [X\bar X]_D + [{\cal W}]_F\,.
 \label{modelchiral}
\end{equation}
The ($Q$ and $S$) supersymmetry transformation laws are
\begin{align}
 \delta X  = & \frac{1}{\sqrt{2}}
 \bar \epsilon P_L\Omega \,, \nonumber\\
 \delta P_L\Omega   = & \frac1{\sqrt{2}} P_L\left(\slashed{\cal D} X +
F\right)\epsilon  +\sqrt{2} X P_L\eta \,,\nonumber\\
\delta F =&\frac1{\sqrt{2}} \bar{\epsilon}\, \slashed{\cal D}
P_L\Omega\,.
\label{chiralconfgauged}
\end{align}
The covariant derivatives are
\begin{align}
  {\cal D}_\mu   X =& \partial _\mu   X -\rmi \, A_\mu
   X -\frac{1}{\sqrt{2}}\bar \psi _\mu P_L\Omega \,, \nonumber\\
   P_L {\cal D}_\mu\Omega  =& P_L\left[ \left( \partial _\mu+\frac14\hat{\omega }_\mu {}^{bc}\gamma _{bc}
  +\ft12\rmi A_\mu\right)\Omega -\frac1{\sqrt{2}} \left(\slashed{\cal D} X +F\right)\psi _\mu  -\sqrt{2} X\phi _\mu \right]
    \nonumber\\
  \phi _\mu  = &- \gamma ^\nu  \left( \partial _{[\mu }-\ft32\rmi
A_{[\mu }\gamma _*+\ft14\hat{\omega } _{[\mu }{}^{ab}\gamma _{ab}\right) \psi
_{\nu ]}+\ft{1}{6}\gamma _\mu \gamma
  ^{\rho \nu }\left( \partial _{\rho }-\ft32\rmi
A_{\rho }\gamma _*+\ft14\hat{\omega } _{\rho }{}^{ab}\gamma _{ab}\right) \psi
_{\nu }\,.
\label{covderscalar}
\end{align}

We present at once the field equation for the scalar field:
\begin{align}
  e^{-1}\frac{\delta S_1}{\delta \bar X }=&\Box^C X +\overline{{\cal W}}''\bar F-\ft12\overline{{\cal W}}'''\bar \Omega  P_R\Omega  \nonumber\\
  &+\frac{1}{\sqrt{2}}\bar \psi_\mu\gamma^\mu\left[\slashed{\cal D}P_L\Omega  +\overline{{\cal W}}''P_R\Omega +\frac{1}{\sqrt{2}}(F +\overline{{\cal W}}')P_R\gamma^\nu\psi_\nu\right]\nonumber\\
  & -\ft12\bar \psi_\mu P_L\psi^\mu\left[ F +\overline{{\cal W}}'\right]\,,
 \label{XL1fe}
\end{align}
Again, several terms have been combined in covariant derivatives (and in the covariant d'Alembertian). The expression in square brackets in the second line is the field equation of $\Omega $, while the one in the last line is the field equation of the auxiliary field.

Thus a covariant set of equivalent equations is
\begin{align}
    & \Box^C X +\overline{{\cal W}}''\bar F-\ft12\overline{{\cal W}}'''\bar \Omega  P_R\Omega \approx 0 \,,\nonumber\\
    & \slashed{\cal D}P_L\Omega  +\overline{{\cal W}}''P_R\Omega \approx 0 \,,\nonumber\\
    & F +\overline{{\cal W}}'\approx 0 \,.
\label{covariantChiralFE}
\end{align}

\subsection{\texorpdfstring{$D=11$}{D=11} supergravity}

A very obvious example is $D=11$ supergravity, already mentioned in the introduction \cite{Cremmer:1978km}. This case can not be treated with the off-shell superspace methods of \cite{Gates:1983nr}. We will concentrate on the gravitino field equation. The relevant part of the action is then
\begin{eqnarray}
S &=& \frac{1}{2\kappa^2}\int \rmd^{11}x\, e \left[R(\omega) \,-\,\bar{\psi}_\mu \gamma^{\mu\nu\rho}D_\nu(\ft12(\omega+\hat{\omega })) \psi_\rho\right.\nonumber\\
&&\left.
\phantom{\frac{1}{2\kappa^2}\int \rmd^{11}} -\frac{\sqrt{2}}{192}
\bar{\psi}_\nu
M^{abcd\nu\rho}
\psi_\rho
( F_{abcd} + \hat{F}_{abcd})+\ldots\right]\,.
\label{fullactD11}
\end{eqnarray}
We use here various quantities. The matrix quantity is
\begin{equation}
  M^{abcd }{}_{\mu \nu }=\gamma^{abcd}{}_{\mu\nu} + 12\gamma^{[ab}e ^c_{[\mu}e ^{d]}_{\nu]}\,.
 \label{defM}
\end{equation}
 For the spin connection, we denote the quantity in (\ref{covomega}) and
\begin{equation}
  \omega_{\mu ab} = \hat{\omega }_{\mu ab} + K^{(5)}_{\mu ab}\,,\qquad \hat{\omega }_{\mu ab} = \omega_{\mu ab}(e)+ K_{\mu ab}\,,\qquad
  K^{(5)}_{\mu ab}= \ft18
\bar\psi_\nu\gamma^{\nu\rho}{}_{\mu ab}\psi_\rho\,.
 \label{omegahat}
\end{equation}
If we want to use the 1.5 order formalism, we should work with $\omega_{\mu ab} $. Indeed the variation of the action w.r.t. the latter is solved by $\omega_{\mu ab} =0$. However, this quantity has supersymmetry transformations proportional to the derivative of the supersymmetry parameter $\epsilon $, while this is not the case for $\hat{\omega }_{\mu ab}$.

The 4-form curvature has the covariant version:
\begin{equation}
  \hat{F}_{\mu\nu\rho\sigma}={F}_{\mu\nu\rho\sigma} +\ft{3}{2}\sqrt{2}\,\bar{\psi}_{[\mu}\gamma_{\nu\rho}\psi_{\sigma]}\,,\qquad {F}_{\mu\nu\rho\sigma}= 4\,\partial_{[\mu}A_{\nu\rho\sigma]}\,.
 \label{hatFF}
\end{equation}
The covariant quantity is in fact $\hat{F}_{abcd}$, which is covariant under the the transformation rules
\begin{eqnarray}
\delta e^a_\mu &=& \ft12 \bar{\epsilon }\gamma^a\psi_\mu\,,\nonumber\\
\delta \psi_\mu &=& D_\mu(\hat\omega) \epsilon +
\frac{\sqrt{2}}{288}
N^{abcd}{}_\mu
\hat{F}_{abcd} \epsilon\,,\qquad N^{abcd}{}_\mu=\gamma^{abcd}{}_\mu\, -\,8e^{[a}_\mu\gamma^{bcd]}\,,\nonumber\\
\delta A_{\mu\nu\rho} &=& -\frac{3\sqrt{2}}{4} \bar{\epsilon }
\gamma_{[\mu\nu}\psi_{\rho]}\,.
\label{fulltrfD11}
\end{eqnarray}
Note the appearance of the covariant quantities $\hat\omega$ and $\hat{F}$ in the transformation of the gravitino. The matrix quantities $M$ and $N$ are related:
\begin{equation}
\gamma ^{\mu \nu \rho }N^{abcd}{}_\mu= -3 M^{abcd\nu \rho }\,.
 \label{relationMN-}
\end{equation}

Due to the difference between the different versions of the spin connection in (\ref{omegahat}), the field equation of (\ref{fullactD11}) for the gravitino looks very complicated:
\begin{align}
  -\kappa ^2e^{-1}\frac{\dl S}{\partial\bar \psi _\mu  } = & \gamma^{\mu\nu\rho}D_\nu(\omega (e)) \psi_\rho
   +\ft18 \left(\gamma^{\mu\nu\rho}\gamma ^{ab}+\gamma ^{ab}\gamma^{\mu\nu\rho}\right)\psi _\rho
  (K+\ft12K^{(5)})_{\nu ab}\nonumber\\ &
  -\ft1{64}\gamma ^{\mu \sigma }{}_{\nu ab}\psi _\sigma \left(\bar \psi _\tau \gamma ^{\tau \nu \rho ab}\psi _\rho
  +2\bar \psi^a\gamma ^\nu \psi ^b\right)  \nonumber\\
  &+\frac{\sqrt{2}}{192}\left[
 M^{abcd \mu \rho }\psi_\rho
( F_{abcd} + \hat{F}_{abcd})+
\frac32\sqrt{2} \gamma _{[bc }\psi _{d ]}\bar \psi _\nu M^{\mu bcd \nu \rho }\psi_\rho\right] \,.
\label{fepsiD11}
\end{align}
This does not look covariant. In this case, since the derivative of the gravitino does not appear in the transformation laws (\ref{fulltrfD11}). The theorems in this paper will prove that therefore (\ref{fepsiD11}) should be covariant.\footnote{After multiplying by a frame field to turn the free index $\mu $ into a flat index} In fact, the covariant expression should be obtained just by dropping the non-covariant terms, and thus is
\begin{equation}
  \gamma^{\mu\nu\rho}\left(D_\nu(\hat{\omega }) \psi_\rho-\frac{\sqrt{2}}{288}  N^{abcd}{}_\nu \psi_\rho \hat{F}_{abcd}\right)\approx 0\,,
 \label{fepsicov}
\end{equation}
A lot of gamma algebra is needed using 3-gravitino Fierz identities (see \cite{D'Auria:1982nx}) to prove that these two expressions are indeed the same.

\section{Basic ingredients}
\label{ss:basic}
\subsection{General result}
We will start by deriving an expression for the variation of the eom under a gauge transformation.
We first use the symbolic DeWitt notation \cite{DeWitt:1967ub} in which the indices $i$ on fields $\{\phi ^i\}$ include the spacetime point (and sums over $i$ include integration over spacetime points).
The invariance of the action under infinitesimal transformations $\delta (\epsilon )\phi ^i$ is then the statement\footnote{The derivative with respect to fermionic fields is taken from the left.}
\begin{equation}
  \delta S= (\delta (\epsilon ) \phi ^i)\frac{\dl S}{\delta \phi ^i} =0\,.
 \label{gaugeinvS}
\end{equation}
The field equations are
\begin{equation}
  S_i\approx 0\,,\qquad S_i\equiv \frac{\dl S}{\delta \phi ^i}\,.
 \label{fieldeqnsT}
\end{equation}
The gauge transformation of the expression in the field equation is thus
\begin{align}
  \delta (\epsilon )S_i=& (\delta (\epsilon ) \phi ^j)\frac{\dl S_i}{\delta \phi ^j}= (\delta (\epsilon ) \phi ^j)\frac{\dl }{\delta \phi ^j}\frac{\dl}{\delta \phi ^i}S\nonumber\\
  =&  \frac{\dl}{\delta \phi ^i}\left((\delta (\epsilon ) \phi ^j)\frac{\dl }{\delta \phi ^j}S\right)-\frac{\dl}{\delta \phi ^i}\left((\delta (\epsilon ) \phi ^j)\right)\frac{\dl }{\delta \phi ^j}S=-\frac{\dl}{\delta \phi ^i}\left((\delta (\epsilon ) \phi ^j)\right)\frac{\dl }{\delta \phi ^j}S \,,
 \label{transfoTi}
\end{align}
using the invariance of the action (\ref{gaugeinvS}). To determine whether a field equation is `covariant', we have to determine whether this transformation contains a derivative on a parameter. This can not be seen from this compact notation. We now have to make the spacetime dependence explicit. We thus write $\phi ^i(x)$ (and thus the index $i$ does not contain the spacetime dependence anymore):
\begin{equation}
  \delta(\epsilon ) S_i(x) =-\int\rmd^4y\,\frac{\dl}{\delta\phi^i(x)}\Big(\delta(\epsilon)\phi^j(y)\Big)\frac{\dl S}{\delta \phi^j(y)}\,.
 \label{delTlocal}
\end{equation}
We assume that \textit{the transformation of fields can depend locally on fields or on derivative of fields}, such that
\begin{equation}
 \frac{\dl}{\delta\phi^i(x)}\Big(\delta(\epsilon)\phi^j(y)\Big)=\frac{\pl \left(\delta(\epsilon)\phi^j(y)\right)}{\partial \phi ^i(y)}\delta (x-y)+
\frac{\pl \left(\delta(\epsilon)\phi^j(y)\right)}{\partial \partial _\mu \phi ^i(y)}\partial ^y_\mu \delta (x-y) \,.
 \label{derdelphilocal}
\end{equation}
Therefore, after a partial integration
\begin{equation}
  \delta(\epsilon ) S_i(x) = -\frac{\partial (\delta(\epsilon)\phi^j(x))}{\partial\phi^i(x)}S_j(x)
+\partial_\mu\left(\frac{\partial(\delta(\epsilon)\phi^j(x))}{\partial\partial_\mu\phi^i(x)}S_j(x)\right)\,.
 \label{delTi}
\end{equation}
$S_i$ is covariant if the right-hand side does not contain spacetime derivatives on the parameters.
This expression shows that the covariance or non-covariance of the field equations depends entirely on the field variations, not for example on the specific form of the Lagrangian.

We will use the terminology `non-covariant terms' to denote terms that explicitly depend on differentiated gauge parameters. A priori the non-covariant terms in (\ref{delTi}) may reside in either term. However it will turn out only the second term is of importance once we've taken care of general coordinate transformations in the following subsection.

\subsection{General coordinate transformations}
\label{ss:gccovfe}
Under general coordinate transformations a field transforms only in itself. E.g. for scalar fields we have
\begin{equation}
  \delta_{\rm gct}(\xi ) \phi = \xi ^\mu \partial _\mu \phi \,.
 \label{gctphi}
\end{equation}
In (\ref{delTi}), the second term thus leads to a derivative of the parameter, and we get
\begin{equation}
  \delta_{\rm gct} (\xi )S_i = \partial _\mu (\xi ^\mu S_i)\,.
 \label{delxiT}
\end{equation}
Using the inverse frame field with
\begin{equation}
  \delta_{\rm gct} (\xi )e^{-1}= \xi ^\mu \partial _\mu e^{-1}-e^{-1}\partial _\mu \xi ^\mu\,,
 \label{delxie-1}
\end{equation}
it is clear that for a scalar field the field equation that is covariant for general coordinate transformations is
\begin{equation}
 T_{\rm cov}= e^{-1}\frac{\delta S}{\delta \phi }\approx 0\,, \qquad \delta_{\rm gct}T_{\rm cov} = \xi ^\mu \partial _\mu  T_{\rm cov}\,.
 \label{covfescalar}
\end{equation}
We already silently applied this in the examples in section \ref{ss:examples}. This covariantization for general covariant transformations with $e^{-1}$ applies for all field equations. However, we have to do more for non-scalar fields under general coordinate transformations.

Fields with a space-time index like $A^\mu $ in section \ref{ss:sME} transform with a product of a field and the differentiated gauge parameter of translations:
\begin{equation}
 \delta _{\rm gct}A^\mu = \xi ^\nu \partial _\nu A^\mu -A^\nu \partial _\nu \xi ^\mu\,.
 \label{gctAmu}
\end{equation}
The non-covariance due to the first term is removed like for the scalar field by including the $e^{-1}$ in the field equation, but the second term in (\ref{gctAmu}) leads according to the first term in (\ref{delTi}) also to derivatives on the parameter $\xi ^\mu $, obtaining the transformation of a vector quantity:
\begin{equation}
  T_\mu \equiv  e^{-1}\frac{\delta S}{\delta A^\mu }\,,\qquad \delta _{\rm gct}T_\mu = \xi ^\nu \partial _\nu T_\mu +(\partial _\mu \xi ^\nu )T_\nu \,.
\label{Tmu}
\end{equation}
It is easy to show that these terms are canceled by multiplying field equations by frame fields to turn them into coordinate scalars as in (\ref{feAmucov}).
We thus define as field equation for vectors
\begin{equation}
  T_a =e^{-1} e_a^\mu\frac{\delta S}{\delta A^\mu }\,.
 \label{Ta}
\end{equation}
It is clear how to generalize this for other `form fields' like antisymmetric gauge tensors $A_{\mu \nu },\ldots $.

Observe that despite the fact that $T_\mu $ is not covariant, it does not transform with a spacetime derivative on parameters of standard gauge transformations. However, it transforms with naked gravitinos under supersymmetry:
\begin{equation}
  \delta T_\mu = \ft12\bar \epsilon \gamma ^a\psi _\mu T_a + e_\mu ^a\delta T_a\,,
 \label{delTmu}
\end{equation}
and the last term transforms as a covariant field.

Since these rules to obtain coordinate scalars are clear and well-known, we will not write this explicit for every field below. We will use the indication $T_i$ to denote the gct-covariant form of the field equation $S_i$, which thus contains e.g. the multiplication with $e^{-1}$. We will further discuss only the `standard gauge transformations', i.e. all gauge transformations excluding general covariant transformations.

\subsection{Standard gauge transformations}
Having treated the general coordinate transformations, we now consider standard gauge transformations.
We discuss the field equations that are already written as a scalar for gct, i.e. as in (\ref{covfescalar}) or  (\ref{Ta}).
The theories that we consider contain the following fields, in general denoted by $\{\phi ^i\}$:
\begin{description}
  \item[Frame field:] $e_\mu ^a$, already used for the gct-covariantization.
  \item[Gauge fields:] fields transforming under standard gauge transformations with a derivative on a parameter, but that term in the transformation does not contain other fields, e.g.
  $\delta A_\mu =\partial _\mu \theta+\ldots  $ or $\delta A_{\mu \nu }= 2\partial_{[\mu }\theta _{\nu ]}$. These will be denoted by $B_\mu{}^A$.
  \item[Covariant fields:] transforming without a derivative on the parameters. These will be denoted by $\Phi_i$.
\end{description}
A general field that fits in either of the above categories will be denoted as in previous subsections, by $\phi_i$.
We got to  (\ref{delTi}) for the transformations of field equations. In the first term there are no derivatives on the parameter for standard gauge transformations. Indeed, $\delta (\epsilon )\phi ^j$ can contain derivatives on the parameter if $\phi ^j$ is a gauge field, but this term is not dependent on other fields.
Hence, it is the second term in (\ref{delTi}) that gives rise to non-covariant terms.

Therefore we can already draw the important conclusion in rule 1:
\begin{tcolorbox}[title=Rule 1]
	\label{rule1}
Assuming that transformation laws depend on fields and their spacetime derivatives, and that the set of fields contains `covariant fields' and `gauge fields', the latter having a transformation law as in (\ref{delBmugeneral}), then (for the gct covariant form of the eom):\vspace{3mm}

	\centerline{The equation of motion of a field $\phi^i$ is NOT covariant}
	\[ \Leftrightarrow \]
	\centerline{There is a field $\phi^j$ that has a derivative on $\phi^i$ in its symmetry variation.}
\end{tcolorbox}
In field theories like the standard model, the gauge transformation laws of the fields do not contain derivatives on other fields, and hence we do not find the situation described above. Therefore all field equations in such theories are covariant. However, in supergravity fields do often transform with a derivative of another field, and therefore the title of this paper contains `supergravity' despite the fact that in this general theorem we did not use any particular property of supersymmetry.

We assumed that the symmetry transformations of fields contain at most first order derivatives on (other) fields. The putative non-covariance of field equations originate from the part of the transformations containing these derivatives, which we can parametrize as\footnote{We always take care of the order of quantities in products such that the formulas are applicable for fermionic fields and parameters. The derivatives with respect to fermionic fields are taken from the left.}
\begin{equation}
 \epsilon ^A H_{Ai}{}^{\mu j}= \frac{ \pl \delta (\epsilon ) \phi ^j}{\partial (\partial _\mu \phi ^i)}\,.
 \label{defHgeneral}
\end{equation}
Thus, the non-covariance of these field equations is due to the term (see (\ref{delTi}))
\begin{equation}
  \delta(\epsilon ) T_i = (\partial _\mu \epsilon ^A)H_{Ai}{}^{\mu j}T_j +\ldots \,.
 \label{delTinoncov}
\end{equation}

\section{Covariant equations of motion}
\label{ss:coveom}

Rule 1 tells us exactly which field equations are already covariant and which aren't. In this section we prove that, under suitable conditions that are satisfied in supergravity due to a dimension counting, covariant field equations can exist that are equivalent to the field equations.
In the next section, we will then show how to construct the covariant field equations, and we show that they are also easy to find. Readers only interested in the result can skip immediately to the box at the end of section \ref{ss:constreom}.
\subsection{A preview}
\label{ss:preview}
If in  (\ref{delTinoncov}), the $H$ and the $T$  in the right-hand side are covariant, the solution would be
\begin{equation}
  \Theta _i = T_i -B_\mu {}^A H_{Ai}{}^{\mu j}T_j=T_i -B_\mu {}^A H_{Ai}{}^{\mu j}\Theta _j\,.
 \label{Thetaisimple}
\end{equation}
We can use this also for obtaining the transformation of field equations
\begin{equation}
  \delta \Theta _i = \left[\delta T_i - \epsilon ^B {\cal M}_{\mu\, B}{}^A H_{Ai}{}^{\mu j}T_j\right]^{\rm cov}\,,
 \label{delThetai}
\end{equation}
The indication $[\cdots ]^{\rm cov}$ means  that the expression is covariantized: all derivatives on matter fields are replaced by covariant derivatives, and gauge fields occur in covariant derivatives of covariant curvatures. Further, ${\cal M}$ is the covariant part of the transformation of the gauge field, see (\ref{modifiedGaugeTrA}) or for more details \cite[(11.69)]{Freedman:2012zz}. For ${\cal N}=1$, $D=4$ the gravitino has such transformations due to the conformal $S$-supersymmetry transformations $\delta \psi _\mu  = -\gamma _\mu \eta $, or the corresponding transformations after gauge fixing to Poincar\'{e} supergravity. For extended supersymmetry or theories in higher dimension, there are more such terms.

What we thus look for is a generalization of this procedure. There are two issues:
\begin{enumerate}
  \item We need an order of the field equations such that when we consider the field equation of $\phi ^i$, the field equations of the $\phi ^j$ that appear in $H_{Ai}{}^{\mu j}$ are already covariantized.
  \item Possibly the $H$ are not covariant. This means that gauge fields appear in $H$ outside of covariant derivatives or curvatures.
\end{enumerate}

As a first example we present a toy model for which we solve the first problem and the second does not appear.

\subsection{Toy model: first example}
 A common case is that the transformations of fields is of the form
\begin{equation}
\label{fieldvar}
\delta(\epsilon)\phi^i=\epsilon^A\, A_A{}^i(\phi ) +\partial_\mu \phi^j\,\epsilon^A\,H_{ Aj}{}^{\mu i}(\phi )  +\partial_\mu\epsilon^A \,R_A{}^{\mu i} \,,
\end{equation}
where $R_A{}^{\mu i}$ is just a non-zero constant for the gauge fields as in (\ref{delBmugeneral}), i.e. for this gauge field $\phi ^i= B_\mu {}^A$ and thus $R_B{}^{\nu}{}_\mu^ A = \delta ^A_B\delta _\mu^\nu $. This setup is thus of the form described at the end of the previous section. The simplification is that the dependence of the transformations on $\partial _\mu \phi $ is linear and $H$ is assumed to be covariant. Hence, (\ref{delTinoncov}) gives the `non-covariant' transformations of the field equations.

Consider the example in section \ref{ss:sME}. The field $\lambda $ does not appear with spacetime derivatives in transformations rules. Therefore we have already
\begin{equation}
  \Theta (\lambda )= T(\lambda )\,,
 \label{Thetalambda}
\end{equation}
and indeed we saw that the (gct corrected) field equation is covariant.

On the other hand $A^\mu $ appears in $F_{ab}$ in the transformation of $\lambda $:
\begin{equation}
  \delta \lambda ^\alpha = -\ft12 (\gamma ^{\mu \nu })^\alpha {}_\beta \epsilon ^\beta \partial _\mu A_\nu \,.
 \label{dellambdaindices}
\end{equation}
Note that we use here explicit spinor indices with the conventions as in \cite[Sec. 3.2.2]{Freedman:2012zz}. In particular, $\lambda ^\alpha $ are the components of $\bar \lambda $.
We put the indices up in order to compare with the abstract notation. The set $\phi ^i$ now contains $\lambda ^\ualpha $ and $A^\umu $. We underline indices when they are values of the index $i$. Further the index for the symmetry $A$ is here the spinor index of the supersymmetry parameter $\epsilon ^\alpha $.

Thus the non-zero component of $H_{Ai}{}^{\mu j}$ corresponding to (\ref{dellambdaindices}) is
\begin{equation}
  H_{\alpha  \unu }{}^{\mu \ubeta  }= -\ft12 (\gamma ^\mu {}_\nu)_{ \alpha  }{}^\beta  \,.
 \label{HsYM}
\end{equation}
This does not contain gauge fields and is covariant.
Using  (\ref{HsYM}) in (\ref{Thetaisimple}), we find
\begin{equation}
  \Theta _\unu = T_\unu +\ft12 \psi _\mu ^\alpha (\gamma ^\mu {}_\nu)_\alpha  {}^\beta  T_\ubeta \,.
 \label{Thetanu}
\end{equation}
Thus the last term in (\ref{feAmucov}) cancels, and one remains with (\ref{covfeSME}) after going to coordinate scalars as explained in section \ref{ss:gccovfe}.

\subsection{Order of eom in supersymmetry.}
\label{ss:ordereomsusy}

We consider here the typical case of supersymmetry, and leave a more abstract and general treatment to section \ref{ss:ConvenientStructure}.
The parameter of supersymmetry has mass dimension\footnote{$[A]$ denotes the mass dimension of $A$.} $[\epsilon]=-1/2$. We will now consider the typical case where $[H_{Ai}{}^{\mu j}]\geq 0$. This includes the usual case where fields have no negative dimension and only zero-dimension fields appear in denominators of transformation rules.

The mass dimensions of the expression (\ref{defHgeneral}) must be consistent, which implies
\begin{equation}
[\phi^j]=[H_{Ai}{}^{\mu j}]+[\phi^i]+[\epsilon^A] +1 > [\phi^i] \mbox{ for }H_{Ai}{}^{\mu j}\neq 0\mbox{ and }[H_{Ai}{}^{\mu j}]\geq 0\,.
\label{dimfieldsH}
\end{equation}
Therefore there is a natural order starting with the highest dimension fields. If $\phi ^i$ are the highest dimension fields then there are no fields $\phi ^j$ satisfying (\ref{dimfieldsH}). Hence
\begin{equation}
  \Theta _i = T_i \qquad \mbox{for the highest dimension fields.}
 \label{Thetahighest}
\end{equation}
We will further order the fields according to decreasing mass dimensions: for fields $\phi^j$ and $\phi^i$ with $[\phi^j]\geq[\phi^i]$ the order should be such that $j<i$.
Hence with this order the non-zero coefficients of $H$ appear only for $j<i$, and $H_{Ai}{}^{\mu j}$ is strictly triangular.
Using this order we will construct the covariant $\Theta _i$ of the form
\begin{equation}
  \Theta _i = T_i - h_i{}^j\Theta _j\,,
 \label{covThetah}
\end{equation}
where the coefficients $h_i{}^j$ are field-dependent and different from zero only for $i>j$, such that this construction can be done perturbatively.
This solves for the supersymmetry case the first problem mentioned at the end of section \ref{ss:preview}. 

\subsection{General case}
\label{ss:coveomgen}

We now show that a solution to the second problem can be found. Our aim is to construct covariant $\Theta _i$ of the general form
\begin{equation}
  \Theta _i = g_i{}^jT_j\,,
 \label{covTheta}
\end{equation}
If the $\Theta_i$ are to be covariant alternatives to the $T_i$, this matrix $g_i{}^j$ should be invertible. This will be further discussed in section~\ref{ss:invert}. We will assume this property for now. The covariance property is the statement that in $\delta \Theta _i$ derivatives of the parameters should cancel. They appear from the non-covariant part of the transformations of $T_i$ as in (\ref{delTinoncov}), and possibly from the non-covariance of $g_i{}^j$. Notice how $\delta T_i$ has terms proportional to the parameter and first derivatives of the parameter, no higher order derivatives as can be verified in (\ref{delTinoncov}). Then $\delta g_i{}^j$ should also depend only on the gauge parameter and its first derivatives. Because of this we will only need to study cancelation of terms proportional to a $\partial_\mu\epsilon^A$ in $ \Theta _i$. To this end we introduce following operator on such a function $f$:
\begin{equation}
D_A^\mu f\equiv\frac{\pl}{\partial \partial_\mu\epsilon^A}\delta(\epsilon)f\,.
\label{DAmydef}
\end{equation}
This operator acts as a derivative, satisfying for example graded distributivity. As an example (\ref{delTinoncov}) can be written as
\begin{equation}
  D_A^\mu T_i = H_{Ai}{}^{\mu j}T_j\,.
 \label{DAmuT}
\end{equation}
Because $\delta\Theta_i$ contains only terms proportional to the parameter and first derivatives of the parameter, the property of covariance can be imposed by requiring
\begin{equation}
D_A^\mu \Theta_i = 0\,,
\label{DAmuTheta}
\end{equation}
and solving this for $g_i{}^j$. To do this we now discuss several ways to evaluate $D_A^\mu $ on functions.

We assumed that the symmetry transformations depend on the fields $\phi ^i$ and their derivatives $\partial _\mu \phi ^i$. We will find solutions where the other functions, such as $g_i{}^j$, also depend on $\{\phi ^i,\,\partial _\mu \phi ^i\}$. If we split the fields in covariant fields we have thus functions of the form
\begin{equation}
  f\left(\Phi ^i,\,\partial _\mu \Phi ^i,\, B_\nu{} ^A,\, r_{\mu \nu }{}^A,\, \partial _{(\mu }B_{\nu )} {} ^A\right)\,, \qquad r_{\mu \nu }{}^A \equiv  2\partial _{[\mu }B_{\nu ]} {} ^A\,.
 \label{ffunction}
\end{equation}
We have split the derivatives of the gauge fields in an antisymmetric and a symmetric part. In transformations of $f$ only the last entry leads to second derivatives on the parameters. If $f$ is a covariant quantity there should thus not be a dependence on the last entry. We will avoid such second derivatives in general by assuming that also other functions, as $g_i{}^j$ do not depend on that symmetric derivative, which is equivalent to assuming $\delta g_i{}^j$ has only terms proportional to the gauge parameter and its first derivatives. We thus have e.g.
\begin{equation}
  \frac{\partial f}{\partial \partial _\mu  B_\nu {}^A}= 2\frac{\partial f}{\partial r_{\mu \nu } {}^A}\,.
 \label{ddBinr}
\end{equation}
In this basis, using the basic transformations (\ref{delcovA}) and (\ref{modifiedGaugeTrA}), we can identify the way in which first derivatives on the symmetry parameters appear in the transformation of a function $f$, leading to
\begin{equation}
  D_A^\mu  f \left(\Phi ^i,\,\partial _\mu \Phi ^i,\, B_\nu{} ^A,\, r_{\mu \nu }{}^A\right)= R_A{}^i\frac{\pl f}{\partial \partial _\mu \Phi ^i} + \frac{\pl f}{\partial B _\mu{}^A}+ 2(B_\nu {}^C f_{CA}{}^B +{\cal M}_{\nu A}{}^B) \frac{\pl f}{\partial r _{\mu\nu }{}^B}\,.
  \label{DefDAmu}
\end{equation}
In a more compact notation, we use all fields $\{\phi ^i\}=\{\Phi ^i,\,B_\mu ^A\}$, and write their symmetry transformations as
\begin{equation}
  \delta (\epsilon ) \phi^i (x)  = \epsilon ^A(x) R_A{}^i( \phi,\partial \phi )(x) +\partial_\mu\epsilon^A(x) \,R_A{}^{\mu i}  \,,
 \label{delcov}
\end{equation}
where $R_A{}^i( \phi,\partial \phi )$ is a local function of fields and its derivatives, and just as in (\ref{fieldvar}) $R_A{}^{\mu i}$ is constant and is non-zero only if $\phi^i (x)$ is a gauge field. This leads to
 \begin{equation}
 D_A^\mu  f(\phi ^i,\,\partial _\mu \phi ^i)= R_A{}^i\frac{\pl f}{\partial \partial _\mu \phi ^i} + \frac{\pl f}{\partial B _\mu{}^A}\,.
 \label{DefDAmucompact}
\end{equation}

We can also introduce a basis of the functions that simplifies the evaluation of  (\ref{DAmydef}).
Therefore, we express the function $f$ using a basis containing the covariant objects (\ref{covariant derivative}) and (\ref{covcurvature}) and $B_\mu ^A$
\begin{equation}
  \tilde f \left(\Phi ^i,\,{\cal D}_\mu \Phi ^i,\, B_\nu{} ^A,\, \hat{R}_{\mu \nu }{}^A\right)= f\left(\Phi ^i,\,\partial _\mu \Phi ^i,\, B_\nu{} ^A,\, r_{\mu \nu }{}^A\right)\,.
 \label{deftildef}
\end{equation}
Then transformations proportional to $\partial _\mu \epsilon ^A$ appear only by the dependence on $B _\mu{}^A$:
\begin{equation}
  D_A^\mu  f= \frac{\pl }{\partial B _\mu{}^A}\tilde f\left(\Phi ^i,\,{\cal D}_\mu \Phi ^i,\, B_\nu{} ^A,\, \hat{R}_{\mu \nu }{}^A\right)\,,
 \label{def2DAmu}
\end{equation}
which shows explicitly that $ D_A^\mu $ is a derivative.

We summarize the different ways to evaluate $D_\mu ^A$ in the following box:
\begin{tcolorbox}[title=Non-covariant part of a function of fields and first derivatives of fields.]
\label{noncovpartsimplefun}
The non covariant part of the variation of a function  $f\left(\Phi ^i,\,\partial _\mu \Phi ^i,\, B_\nu{} ^A,\, r_{\mu \nu }{}^A\right)$ is given by
\begin{align}
 D_A^\mu  f\left(\Phi ^i,\,\partial _\mu \Phi ^i,\, B_\nu{} ^A,\, r_{\mu \nu }{}^A\right) &=  R_A{}^i\frac{\pl f}{\partial \partial _\mu \phi ^i} + \frac{\pl f}{\partial B _\mu{}^A}\,,\nonumber\\
 &=R_A{}^i\frac{\pl f}{\partial \partial _\mu \Phi ^i} + \frac{\pl f}{\partial B _\mu{}^A}+ 2(B_\nu {}^C f_{CA}{}^B +{\cal M}_{\nu A}{}^B) \frac{\pl f}{\partial r _{\mu\nu }{}^B}\,.
 \label{DefDAmuorig}
\end{align}
or in a covariant basis, using (\ref{deftildef}),
\begin{equation}
D_A^\mu  f\left(\Phi ^i,\,\partial _\mu \Phi ^i,\, B_\nu{} ^A,\, r_{\mu \nu }{}^A\right) = \frac{\pl }{\partial B _\mu{}^A}\tilde f\left(\Phi ^i,\,{\cal D}_\mu \Phi ^i,\, B_\nu{} ^A,\, \hat{R}_{\mu \nu }{}^A\right)\,.
\end{equation}
\end{tcolorbox}

The requirement of covariance of $\Theta _i$ is thus\footnote{Sign factors appear in order to move $D_A^\mu $ to the left: $(-)^{A(i+j)}$. This is a minus sign if the symmetry $A$ is fermionic and the fields indexed by $i$ and $j$ have opposite statistics.}
\begin{equation}
  0= D_A^\mu \Theta _i=D_A^\mu\left(g_i{}^jT_j\right)= \left(D_A^\mu g_i{}^j + (-)^{A(i+j)}g_i{}^j H_{Aj}{}^{\mu k}\right)T_k\,,
 \label{DTheta0}
\end{equation}
where we used (\ref{DAmuT}).
If this has to vanish for generic $T_k$, we get a differential equation.
\begin{tcolorbox}[title=Constructing covariant field equations.]
	\label{covfe}
	A covariant field equation can be found if one can solve
\begin{equation}
 D^\mu _A g_i{}^k= - (-)^{A(i+j)}  g_i{}^j H_{Aj}{}^{\mu k}=-(-)^{Ai}g_i{}^j H_{jA}{}^{\mu k}\,,
 \label{diffeqg}
\end{equation}
for the invertible matrices $g_i{}^j$. They determine the covariant $\Theta _i$ in (\ref{covTheta}).
\end{tcolorbox}

In order to restrict the number of sign factors we used in the last expression in (\ref{diffeqg}) a different notation for the function $H$ defined in (\ref{defHgeneral})
\begin{equation}
  H_{jA}{}^{\mu i}= (-)^{jA}H_{Aj}{}^{\mu i}= \frac{\pl R_A{}^i( \phi,\partial \phi ) }{\partial \partial _\mu \phi ^j}\,.
 \label{defHgeneralb}
\end{equation}
\medskip
We can derive an integrability condition for (\ref{diffeqg}) from the super-commutativity of the $D^\mu_A$:
\begin{equation}
D^\mu_A D^\mu_B - (-)^{AB}D^\mu_B D^\mu_A=0\,,
\label{commutatingDmuA}
\end{equation}
which can readily be verified from (\ref{DAmydef}) or (\ref{def2DAmu}). Using (\ref{diffeqg}) and the invertibility of $g_i{}^j$ we can rewrite the left-hand side of (\ref{commutatingDmuA}), and require that it still vanishes. This gives us the condition (more details of the derivation will be given below):
\begin{equation}
I_{iBA}{}^{\nu \mu k}-(-)^{AB}I_{iAB}{}^{\mu \nu k}=0\,,
\label{integrability}
\end{equation}
where
\begin{equation}
I_{iBA}{}^{\nu \mu k}\equiv   H_{iB}{}^{\nu j}H_{jA}{}^{\mu k} - (-)^{iB}D^\nu _B H_{iA}{}^{\mu k} \,.
 \label{defI}
\end{equation}
There is an aesthetic way to arrive at this condition.
We can interpret (\ref{diffeqg}) as an algebraic equation for covariant derivatives on the space of matrices that are functions of fields.\footnote{Readers that get confused by the sign factors may at first put all these statistics factors to $0$ (bosonic), and then notice that the sign factors only appear when the order of fermionic indices has been modified.}
We define for a super-matrix where the coefficients $f_i{}^j$ have statistics $(-)^{f+i+j}$
\begin{equation}
L^\mu_A(f)_i{}^j\equiv D^\mu_A f_i{}^k + (-)^{A(f+i+j)}  f_i{}^j H_{Aj}{}^{\mu k}= D^\mu_A f_i{}^k + (-)^{A(f+i)}  f_i{}^j H_{jA}{}^{\mu k}\,.
\label{defoperatorL}
\end{equation}
The right-hand side is for each $(\mu A)$ a new matrix with elements with statistics $(-)^{A+i+j}$.
E.g., the components of $g_{i}{}^j$ have statistics $(-)^{i+j}$, which is thus a `bosonic' supermatrix and the equation to solve, (\ref{diffeqg}), can be written as
\begin{equation}
L^\mu_A(g)=0\,.
\label{LmuAg0}
\end{equation}
One might consider this equation as the statement of covariant constancy of the matrices $g$  for the derivatives $D^\mu _A $, interpreting the $H_{Ai}{}^{\mu j}$ as one-form (indices $(A\mu)$) matrices (indices $i,j$). Covariantly constant objects can only exist in spaces without curvature, we should thus verify this is indeed the case.
The operators $L^\mu_A$ can define a curvature:\footnote{We assume here that $f_i{}^j$ has statistics $(-)^{i+j}$, as is the case for $g_i{}^j$ and thus $L_A^\mu f_i{}^j$ has statistics $(-)^{A+i+j}$. Taking the opposite statistics for $f$  leads to minus signs that cancel when the two $L$ operators are moved to the right of $f$.}
\begin{align}
L^\nu_B\, L^\mu_A(f)=&L^\nu_B\left( D^\mu_A f_i{}^k + (-)^{Ai}  f_i{}^j H_{jA}{}^{\mu k}\right)\nonumber\\
=&D^\nu_B D^\mu_A f_i{}^k + (-)^{Ai} D^\nu_B\left( f_i{}^j H_{jA}{}^{\mu k}\right)+(-)^{B(A+i)}\left( D^\mu_A f_i{}^\ell + (-)^{Ai}  f_i{}^j H_{jA}{}^{\mu \ell}\right)H_{\ell B}{}^{\nu k}\,.
\label{LLBA}
\end{align}
Therefore
\begin{equation}
L^\nu_B\, L^\mu_A(f)-(-)^{AB} L^\mu_A\, L^\nu_B(f)=-(-)^{(A+B)i} f_i{}^j\left[I_{jBA}{}^{\nu \mu k} -(-)^{AB}I_{jAB}{}^{ \mu\nu k}\right]\,,
 \label{curvatureL}
\end{equation}
and so we find that the condition of vanishing of the curvature and the integrability condition derived in (\ref{integrability}) are equivalent, as of course they should be.\medskip

We now show that this integrability condition is indeed satisfied using two pieces of information. Firstly we found a general equality (\ref{HHrewritten}) for $D^\nu _B H_{iA}{}^{\mu k}$ in appendix \ref{app:covariant} using that covariant fields have covariant field variations and that the variation of a gauge field has a well defined non-covariant part. When we use this in (\ref{defI}), we find
\begin{equation}
  I_{iBA}{}^{\nu \mu k}= H_{iB}{}^{\nu j}H_{jA}{}^{\mu k} + H_{iB}{}^{\mu j}H_{jA}{}^{\nu k}\,.
 \label{simplifiedintegr}
\end{equation}
Hence, this expression is now symmetric in $(\mu \nu )$.

A second equation can be obtained from a consistency requirement on the commutator of transformations.
We consider the commutator relation of two transformations on $\phi ^k$. Since the transformations contain only first order derivatives, double spacetime derivative on the fields can only be obtained in the part of the transformation where each $\delta (\epsilon )$ provides a derivative:
\begin{align}
 \left[ \delta (\epsilon _1),\,\delta (\epsilon _2) \right]\delta \phi ^k &=
  \ldots +\epsilon _2^A\partial _\mu\left(\epsilon _1^B R_B{}^j \right)\frac{\pl R_A{}^k}{\partial \partial _\mu \phi ^j}-(1\leftrightarrow 2)\nonumber\\
  &=\ldots +\epsilon _2^A\epsilon _1^B\partial _\mu \partial _\nu \phi ^i\frac{\pl  R_B{}^j }{\partial \partial _\nu \phi ^i}\frac{\partial R_A{}^k}{\pl \partial _\mu \phi ^j}-(1\leftrightarrow 2)\nonumber\\
 &= \ldots +\epsilon _2^A\epsilon _1^B\partial _\mu \partial _\nu \phi ^i H_{iB}{}^{\nu j}H_{jA}{}^{\mu k}-(1\leftrightarrow 2)\nonumber\\
  &=\ldots +\ft12\epsilon _2^A\epsilon _1^B (\partial _\mu \partial _\nu \phi ^i)I_{iBA}{}^{\nu \mu k}-(1\leftrightarrow 2)\,.
  \label{commconstr1}
\end{align}
No second derivatives should appear in this commutator if either
\begin{enumerate}
  \item There is a closed algebra with structure functions that do not depend on second derivatives.
  \item The algebra is closed modulo field equations that are first order in spacetime derivatives (which is usually the case for fermion eom's.).
\end{enumerate}
In these cases, which are very common in supergravity, the absence of the second derivatives implies that the integrability condition (\ref{integrability}) is satisfied. This implies that a solution to (\ref{diffeqg}) is possible. If there is then an invertible $g_i{}^j$ that solves (\ref{diffeqg}) we can conclude that a set of equivalent, covariant field equations $\Theta_i=0$ exists. We start the following section by determining the necessary property for the $\Theta_i$ to be equivalent to the $T_i$. Or in other words, the necessary properties for $g_i{}^j$ to be invertible. We then explicitly construct a solution to (\ref{diffeqg}) and show that this property is satisfied.

\section{Structure of convenient supergravity theories}
\label{ss:ConvenientStructure}

In the previous section we could use the dimension of fields to find an order in which the field equations are covariantized. In the first part of this section we will considering this ordering problem in general and show when there can be problems in this part of the procedure. In the second part we will explicitly construct the covariant equations.

\subsection{Invertibility condition}
\label{ss:invert}

We now describe a numbering of the fields giving the order in which their field equations will be covariantized. We assume that there are $n_0$ fields whose field equations $T_i$ are already covariant. These are thus the fields that do not appear with derivatives in any transformation law, and therefore $\delta (\epsilon )T_i$ has no derivatives on gauge parameters. We number these as $T_i$ with $i=1,\ldots , n_0$. The fields for which $(\delta(\epsilon)T_j)|_{T_1=T_2 =\dots = T_{n_0}=0}$ has no derivatives on gauge parameters, are numbered from $n_0+1$ to $n_1$. The fields for which $(\delta(\epsilon)T_j)|_{T_1=T_2 =\dots = T_{n_1}=0}$ have no derivatives on gauge parameters, are numbered from $n_1+1$ to $n_2$, and so on. If all fields can be renumbered in this way, the on-shell field configurations can be described as
\begin{equation}
\begin{split}
\bigg\{ \phi^j \bigg| T_i=0 \bigg\}=&\bigg\{ \phi^j \bigg| T_1=0 \,; \, T_2=0 \,;\, \dots\, ;\, T_{n_0}=0 \bigg\} \\
&\cap\bigg\{ \phi^j \bigg| T_{n_0+1}\big|_{T_1=\dots=T_{n_0}=0}=0\, ; \, \dots\, ; \,T_{n_1}\big|_{T_1=\dots=T_{n_0}=0}=0 \bigg\} \\
&\cap\bigg\{ \phi^j \bigg| T_{n_1+1}\big|_{T_{1}=\dots=T_{n_1}=0}=0\, ; \, \dots \,;\, T_{n_2}\big|_{T_{1}=\dots=T_{n_1}=0}=0 \bigg\} \\
&\cap \dots\,.
\end{split}
\label{orderedT}
\end{equation}
Here and further we use
\begin{equation}
  T_i|_{T_j=0}= \{T_i\mbox{ restricted to field configurations that satisfy }T_j=0\}\,.
 \label{Ti|Tj}
\end{equation}
If this renumbering can be done, we will look for the covariant expressions $\Theta_i$ in this order, such that for all $i$, they are equivalent to the $T_i$ modulo the already considered field equations:
\begin{equation}
\Theta_i\big|_{T_{1}=\dots=T_{i-1}=0}=T_{i}\big|_{T_{1}=\dots=T_{i-1}=0}\,.
\label{thetacondition}
\end{equation}

Note that it's impossible to find a $\Theta_i$ that is covariant and satisfies this relation if $(\delta(\epsilon)T_i)\big|_{T_{1}=\dots=T_{i-1}=0}$ has derivatives on gauge parameters.

The equality (\ref{orderedT}) states that for every $i$ we have to consider the field equations only modulo the previous ones. Then, if we were able to find covariant $\Theta_i$, we can apply (\ref{thetacondition})
\begin{equation}
\begin{split}
\bigg\{ \phi^j \bigg| T_i=0 \bigg\}=&\bigg\{ \phi^j \bigg| T_i\big|_{T_1=\dots=T_{i-1}=0}=0\bigg\} \\
=&\bigg\{ \phi^j \bigg| \Theta_i\big|_{T_1=\dots=T_{i-1}=0}=0\bigg\} \\
=&\bigg\{ \phi^j \bigg| \Theta_i=0\bigg\} \,,
\end{split}
\end{equation}
where the last equality holds because $\Theta_1=T_1\,$, and further by induction.\footnote{As an illustration we work this out for a system with three fields:
	\begin{equation}
	\begin{split}
	&\bigg\{ \phi^j \bigg| T_1=0;\Theta_2\big|_{T_1=0}=0;\Theta_3\big|_{T_1=T_{2}=0}=0 \bigg\} = \bigg\{ \phi^j \bigg| T_1=0;\Theta_2=0;\Theta_3\big|_{T_1=T_{2}=0}=0 \bigg\}\\
	=&\bigg\{ \phi^j \bigg| T_1=0;\Theta_2=0;\Theta_3\big|_{T_1=T_{2}|_{T_1=0}=0}=0 \bigg\} = \bigg\{ \phi^j \bigg| T_1=0;\Theta_2=0;\Theta_3\big|_{T_1=\Theta_{2}|_{T_1=0}=0}=0 \bigg\}\\
	=&\bigg\{ \phi^j \bigg| T_1=0;\Theta_2=0;\Theta_3\big|_{T_1=\Theta_{2}=0}=0 \bigg\}= \bigg\{ \phi^j \bigg| T_1=0;\Theta_2=0;\Theta_3=0 \bigg\}\,.\\
	\end{split}
	\end{equation}}
Thus the $\Theta_i$ have the same solutions as the original $T_i$.

Notice how in the basis of renumbered fields all matrices $H_{Ai}{}^{\mu j}$ are strictly triangular (in $i$ and $j$).
Indeed,  (\ref{delTinoncov}) says that in order that $T_i|_{T_{k<i}=0 }$ should be covariant, we should have $H_{Ai}{}^{\mu j}=0$ for $j\geq i$. This then implies by the definition of $H$ in (\ref{defHgeneral}) that a field $\phi^j$ cannot transform into a derivative of a field $\phi^i$ with $j\geq i$.

How could this renumbering fail? If the transformation of a field contains the derivative of the same field ($i=j$ above), we cannot continue here. This is the case for general coordinate transformations, but we treated these already in section \ref{ss:gccovfe}.  Further, this can only happen if there is a sort of `loop' in the field variations of the following form: $T_a$ transforms covariantly on field configurations satisfying $T_b=0$, and $T_b$ transforms covariantly on field configurations satisfying $T_a=0$. Or in other words $\delta\phi^a$ and $\delta\phi^b$ contain terms with $\partial_\mu\phi^b$ and $\partial_\mu\phi^a$ respectively. It is clear that renumbering these fields as described above should result in new indices for the fields $a^\prime$ and $b^\prime$ that must satisfy both $b^\prime>a^\prime$ and $a^\prime>b^\prime$. The renumbering can therefore not be done.
This poses a problem because
\begin{equation}
\bigg\{ \phi^j \bigg| T_a=0; T_b=0 \bigg\}\not= \bigg\{ \phi^j \bigg| T_a\big|_{T_b=0}=0; T_b\big|_{T_a=0}=0 \bigg\} \,.
\end{equation}
So though we may find covariant $\Theta_a$ and $\Theta_b$ satisfying
\begin{equation}
\Theta_a\big|_{T_{b}=0}=T_{a}\big|_{T_{b}=0}\qquad \text{and}\qquad \Theta_b\big|_{T_{a}=0}=T_{b}\big|_{T_{a}=0}\,,
\end{equation}
they will not (necessarily) satisfy
\begin{equation}
\bigg\{ \phi^j \bigg| T_a=0; T_b=0 \bigg\}=\bigg\{ \phi^j \bigg| \Theta_a=0; \Theta_b=0 \bigg\} \,.
\label{nottrue}
\end{equation}
What would however be true is the following
\begin{equation}
\bigg\{ \phi^j \bigg| T_a=0; T_b=0 \bigg\}=\bigg\{ \phi^j \bigg| T_a=0; \Theta_b=0 \bigg\} \,.
\label{true}
\end{equation}
So in the case of this specific loop structure, we could replace a system of two non-covariant eom by a system of one covariant and one non-covariant eom.\footnote{As a side remark we mention that we expect that in general the minimal number of non-covariant eom, is exactly the number of `independent loops' in the field variations. We have not worked this out in more detail}
This type of loop structure can be traced back to a non-triangularity property of $H_{Ai}{}^{\mu j}$: it is a fundamental property of the field variations.

We conclude with this box:
\begin{tcolorbox}[title=Invertibility of $g_i{}^j$]
\label{rule3}
\centerline{All $H_{Ai}{}^{\mu j}$ are simultaneously strictly triangularizable in $i$ and $j$ by means of a }
\centerline{permutation transformation on the fields}
\begin{equation} \Rightarrow \end{equation}
\centerline{$g_i{}^j$ can be assumed to be invertible.}
\end{tcolorbox}

This property of $H_{Ai}{}^{\mu j}$ can be easily verified for supersymmetry as discussed in section \ref{ss:ordereomsusy}, making use of the dimension of the fields.

\begin{tcolorbox}
If all $H_{Ai}{}^{\mu j}$ have mass dimension $\geq 0$ in supersymmetry, $g_i{}^j$ can be assumed to be invertible.
\end{tcolorbox}

\subsection{Constructing the covariant equations of motion}
\label{ss:constreom}
We will find now a general solution to the condition (\ref{LmuAg0}).
We found a first order solution in (\ref{Thetaisimple}) for the case that $H$ is covariant. Observe that the definition (\ref{defoperatorL}) implies for $f=\unity $.
\begin{equation}
    L_A^\mu (\unity)_i{}^j = (-)^{Ai} H_{iA}{}^{\mu j}=H_{Ai}{}^{\mu j}\,.
 \label{Lunity}
\end{equation}
Therefore the solution  (\ref{Thetaisimple}) can be written as
\begin{equation}
  \Theta _i = T_i - B_\mu {}^A L_A^\mu (\unity)_i{}^j T_j\,,\qquad \mbox{i.e.}\qquad  g_1= - B_\mu {}^A L_A^\mu (\unity)\,.
 \label{ThetaSol1}
\end{equation}
In that case, $H$ was covariant, and thus $D_B^\mu H_{Ai}{}^{\mu j}=0$ and with  (\ref{HHrewritten}) we find that also
\begin{equation}
  L_B^\nu L_A^\mu (\unity)=0\,.
 \label{simpleLL0}
\end{equation}

If this is not so, we will need higher order terms in the gauge fields $B_\mu {}^A$ in order to build the covariant field equation $\Theta _i$. We will therefore solve (\ref{diffeqg}) perturbatively:
\begin{equation}
  g_i{}^j =\sum_{\chi =0} (g_\chi )_i{}^j\,,\qquad (g_0)_i{}^j = \delta _i{}^j\,,
 \label{perturbg}
\end{equation}
and thus find a covariant field equation
\begin{equation}
\Theta_i= T_i + \sum_{\chi =1} (g_\chi )_i{}^jT_j\,,
\label{theta_expansie}
\end{equation}
where $\chi $ counts the power in the gauge fields. We consider now how $L_A^\mu $ acts on products of gauge fields $B_\nu ^B$ and other matrices $f_B^\nu {}_i{}^j$. The definition implies
\begin{equation}
  L_A^\mu\left(B_\nu ^B\,f_B^\nu {}_i{}^j\right)= f_A^\mu {}_i{}^j + (-)^{AB}B_\nu ^B L_A^\mu f_B^\nu {}_i{}^j\,.
 \label{LonBf}
\end{equation}
We can therefore make following ansatz for a perturbative solution of  (\ref{LmuAg0}):
\begin{equation}
\label{sol}
g_\chi=\frac{(-1)^\chi}{\chi!}B_{\mu_1}^{B_1}B_{\mu_2}^{B_2}\cdots B_{\mu_\chi}^{B_\chi} L^{\mu_\chi}_{B_\chi}\cdots L^{\mu_2}_{B_2}L^{\mu_1}_{B_1}\unity\,,
\end{equation}
where it is understood that the $g_0=\unity$, and $g_1$ is as in (\ref{ThetaSol1}). Indeed, repetitive use of (\ref{LonBf}) gives
\begin{align}
  L_A^\mu(g_\chi )= & \frac{(-1)^\chi}{(\chi-1)!}B_{\mu_2}^{B_2}\cdots B_{\mu_\chi}^{B_\chi} L^{\mu_\chi}_{B_\chi}\cdots L^{\mu_2}_{B_2}L^{\mu}_{A}\unity \nonumber\\
    & +(-)^{A(B_1+\ldots +B_\chi )} \frac{(-1)^\chi}{\chi!}B_{\mu_1}^{B_1}B_{\mu_2}^{B_2}\cdots B_{\mu_\chi}^{B_\chi} L_A^\mu L^{\mu_\chi}_{B_\chi}\cdots L^{\mu_2}_{B_2}L^{\mu_1}_{B_1}\unity\,.
\label{Lonchi}
\end{align}
Using on the last term the graded commutation relation of the operators $L$ as in (\ref{curvatureL}) with (\ref{integrability}), we can write this as
\begin{equation}
  L_A^\mu(g_\chi )= -g_{\chi -1}L^{\mu}_{A}\unity + g_\chi L^{\mu}_{A}\unity\,.
 \label{Lgchichim1}
\end{equation}
Summing over $\chi $ readily proves that  (\ref{perturbg}) with (\ref{sol}) solves the requirement (\ref{LmuAg0}) $L_A^\mu g=0$.

Looking at (\ref{theta_expansie}), we can make two important observations. Firstly it is clear from (\ref{sol}) that all $(g_\chi )_i{}^j$ are proportional to at least $\chi $ free gauge fields. Therefore the covariant expression $\Theta _i$ in (\ref{theta_expansie}) is obtained from $T_i$ by covariantizing all non-covariant expressions appearing in this field equation. This means removing all terms with free gauge fields and replacing all derivatives with covariant derivatives and curvatures.

Secondly we can verify that $\Theta_i=0$ has the same solutions as $T_i=0$ by demonstrating that (\ref{thetacondition}) holds. Assume we have renumbered our fields as described in section \ref{ss:invert} and that all $H_{Ai}{}^{\mu j}$ are strictly triangular (in $i$ and $j$). Then (\ref{sol}) tells us that $(g-\unity )_i{}^j$ is also strictly triangular. It follows that
\begin{equation}
\Theta_i\big|_{T_1=T_2=\dots=T_{i-1}=0}= \Big( T_i + \sum_{\chi>0} (g_\chi)_i{}^jT_j\Big)\Big|_{T_1=T_2=\dots=T_{i-1}=0} =  T_i\big|_{T_1=T_2=\dots=T_{i-1}=0}\,.
\end{equation}
Due to the analysis of section \ref{ss:invert} this finally shows $g_i{}^j$ is invertible.
\bigskip

We summarize this in the following box, which is our main, self-contained, result:

\begin{tcolorbox}[title=Rule 2]
If the non-closure functions depend only on fields and first order derivatives of fields and the $H_{Ai}{}^{\mu j}$ can be made strictly triangular\footnote{This is true for supersymmetry if (but not only if) $[H_{Ai}{}^{\mu j}]\geq 0$ for all $H_{Ai}{}^{\mu j}$.}, we can find covariant $\Theta_i=0$ with the same solutions as the eom. One can find these $\Theta_i$ by assuming covariance while calculating $\delta S / \delta \phi^i$. \footnote{One should also multiply with $e^{-1}$ and contract with the correct number of frame fields to covariantize w.r.t. g.c.t.}
\end{tcolorbox}

We show how such a solution is obtained in the example of section \ref{ss:exchiral} in appendix \ref{app:exchiralg}.

\section{Conclusions}
\label{ss:conclusions}

We have considered a rather general type of supergravity theories as summarized at the end of section \ref{ss:basic}.
We first derived a general result for the variation of field equations in (\ref{delTi}). We used this to study the suspected property that field equations should be covariant modulo other field equations. In sections \ref{ss:coveom} and \ref{ss:invert} we found the conditions for the field equations to indeed have this property. We found that (\ref{integrability}) should be satisfied and the $H$'s should be triangulable in sections \ref{ss:coveom} and \ref{ss:invert} respectively. We then found that these conditions are satisfied if (but not only if) the non-closure functions depend only on fields and first derivatives of fields\footnote{Or in other words, if the non-closure functions are all proportional to fermionic field equations that are first order in derivatives.} and if there is no field whose field variation contains a differentiated field of greater mass dimension. These two conditions hold true for most imaginable supergravities, and so in most cases the field equations are covariant modulo other field equations.

In section \ref{ss:constreom} we found that for supergravities satisfying the above described criterion have a set of covariant equations with the same solutions as the equations of motion. These can be easily found by covariantizing the field equations. Reconstructing the equations of motion from these covariant expressions is done by (\ref{perturbg}) and (\ref{sol}). An example of this procedure can be found in appendix \ref{app:exchiralg}.

Supergravity actions are often very long expressions, and tricks to simplify the expressions for the equations of motion are welcome. One of these important tricks is covariance, which has been used by many authors. We have proven theorems to support the covariance arguments. These can be used in many common supergravity actions. In the 40 years of supergravity, the structure of supergravity actions have become more and more general. It is therefore important to know what the conditions are under which these covariance tricks can be applied. This paper contributed to this knowledge. We expect that in the future when more general actions and transformations will be considered, these theorems can still be generalized.

Our methods are also useful for the calculation of the supersymmetry transformations of field equations. The results of this paper show that covariant field equations can be used, for which the calculation of the transformations is easier.  This is applied in \cite{Ferrara:2017yhz}, where this method is summarized in appendix D, to construct current multiplets. For these properties the first order corrections $(g_1)$ in  (\ref{ThetaSol1}) are needed, as we have seen also in (\ref{delThetai}).

\medskip
\section*{Acknowledgments.}

\noindent

We are grateful to Gabriele Tartaglino-Mazzucchelli and Magnus Tournoy for discussions on these results.
This work is supported
in part by the Interuniversity Attraction Poles Programme
initiated by the Belgian Science Policy (P7/37),
and has been supported in part by COST Action MP1210 `The
String Theory Universe'. Further support came from the EC through the grant QUTE. This paper is based on research done in the context of the master thesis of B.V. in KU Leuven, 2016.
\newpage
\appendix
\section{Covariant quantities}
\label{app:covariant}

In this appendix we summarize some statements on covariant quantities developed in \cite{VanProeyen:1983wk,Bergshoeff:1986mz,Freedman:2012zz}. The setup that we consider is a theory with gauge symmetries, which we split in general coordinate transformations (gct), with parameter $\xi ^\mu  (x)$ and a set of other gauge transformations, labeled with an index $A$, and with parameters $\epsilon ^A(x)$. The latter are denoted as `standard gauge symmetries'. In general, we assume that gauge transformations of fields contain at most first derivatives on parameters and other fields.

Let us start with a definition:
\begin{tcolorbox}[title=Definition of a covariant quantity]
	\label{covariant}
	A covariant quantity is a local function that transforms under all local symmetries with no spacetime derivatives of a transformation parameter.
\end{tcolorbox}

Considering first the gct. The definition already implies that an object with a local index $\mu $ cannot be a covariant quantity, since in its gct, e.g. $\delta_{\rm gct} A_\mu =\xi ^\nu \partial _\nu A_\mu +(\partial _\mu\xi ^\nu) A_\nu $, there is a term with a derivative acting on the parameter $\xi ^\nu $. Thus, covariant quantities should be coordinate scalars.

\emph{The set of fields} in the theories that we consider contains the frame field $e_\mu {}^a(x)$, which is the gauge field of gct, gauge fields $B_\mu {}^A(x)$ and covariant fields $\Phi ^i(x)$. We do not consider here $p$-form gauge fields and corresponding $(p-1)$-form symmetries, though we believe that theories with $p$-forms also satisfy the theorems derived in this paper. However, their inclusion in the formulation of the general structures would make this formalism more heavy.

The gauge fields $B_\mu {}^A(x)$ are defined as the fields that transform as
\begin{equation}
  \delta B_\mu {}^A= \partial _\mu \epsilon ^A + \mbox{ terms without spacetime derivatives on the parameters.}
 \label{delBmugeneral}
\end{equation}
Often, the set of fields $\{\Phi ^i,\,e_\mu {}^a,\,B_\mu {}^A\}$ is an over-complete basis. Indeed, some gauge symmetries are gauged by composites of other fields. Examples are the Lorentz transformations, which are gauged by the spin connection $\omega _\mu {}^{ab}(e,\ldots  )$, the $S$-supersymmetry and the special conformal transformations in the superconformal group, which are also gauged by composite fields.

Clearly, a gauge field is not a covariant quantity. We assume that all other fields $\{\Phi ^i\}$ in the basis of fields are covariant fields.
This assumption is reasonable. Indeed, if we would have other fields that are not covariant, which means that they would transform as
\begin{equation}
  \delta (\epsilon )\Phi ^i = (\partial _\mu \epsilon ^A) S_A{}^{\mu i}(\phi )+\ldots\,,
 \label{noncovfield}
\end{equation}
then we could consider the field
\begin{equation}
  \Phi '^i = \Phi ^i - B_\mu {} ^A S_A{}^{\mu i}(\phi )
 \label{phiprimecov}
\end{equation}
in our basis of sets of fields. At least if $R_A{}^{\mu i}(\phi )$ does not contain gauge fields, this field $\Phi '^i$ is covariant. If $R_A{}^{\mu i}(\phi )$ would contain gauge fields, then a similar procedure is possible modifying the factor in the second term of (\ref{phiprimecov}) or adding more correction terms.
\medskip

We can define \emph{covariant derivatives} of covariant fields, and \emph{covariant curvatures} of gauge fields.
The transformation of a covariant quantity $\Phi (x)$, under the standard gauge symmetries is of the form
\begin{equation}
  \delta (\epsilon ) \Phi^i (x)  = \epsilon ^A(x) R_A{}^i( \phi,\partial \phi )(x) \,,
 \label{delcovA}
\end{equation}
where $R_A{}^i( \phi,\partial \phi )$ is a local function of fields and its derivatives.
We can define a covariant derivative of a covariant quantity as follows:
\begin{tcolorbox}[title=Definition of the covariant derivative]
	If $\Phi $ is a covariant quantity, its covariant derivative is given by
	\begin{equation}
	\label{covariant derivative}
{\cal D}_a \Phi  =e_a^\mu 	\mathcal{D}_\mu \Phi \,,\qquad 	\mathcal{D}_\mu \Phi = \left(\partial_\mu - \delta (B_\mu)\right)\Phi = \partial_\mu\Phi(x)-B_\mu{}^A(x)R_A (x) \,.
	\end{equation}
	where $\delta (B_\mu)\Phi $ should be interpreted as the rule (\ref{delcovA}) with the gauge parameters $\epsilon ^A$ replaced by $B_\mu{}^A$.
\end{tcolorbox}

Transformations of gauge fields in supersymmetry can have other terms than in Yang-Mills theory:
\begin{equation}
  \delta (\epsilon)B _\mu {}^A= \partial_\mu\epsilon^A +\epsilon^C B _\mu{}^B
f_{BC}{}^A+ \epsilon^B {\cal M}_{\mu B}{}^A\,.
\label{modifiedGaugeTrA}
\end{equation}
Here we assume that ${\cal M}_{a B}{}^A= e_a{}^\mu {\cal M}_{\mu B}{}^A$ is a covariant quantity. If it would not, then it would contain gauge fields, and we could include it in the term with $f_{BC}{}^A$.
Such ${\cal M}$-terms may originate from a nonzero commutator between  translations and symmetry $B$ to symmetry $A$.  If translations would have been included in the sum over $B$ in the second term, this would lead to a term proportional to $e_\mu{}^b$. The latter is not included in the set $\{B_\mu {}^B\}$, and thus this part of the transformation of $B _\mu {}^A$ is included in ${\cal M}_{\mu B}{}^A$.  Another frequent case is that there are non-gauge fields in the supersymmetry-multiplet of the gauge fields. This is the case in our example of section \ref{ss:sME} since the transformation of the gauge field $A_\mu $ contains the gaugino $\lambda $, see (\ref{gauge multiplet variations}) and (\ref{MmuME}).

With these definitions we can construct a covariant curvature $\widehat{R}_{ab}{}^A$:

\begin{tcolorbox}[title=Definition of covariant curvatures]
 	The covariant curvature $\widehat{R}_{ab}{}^A$  of $B^A_\mu$ is defined as
 	\begin{equation}
\widehat{R}_{ab}{}^A= e_a {}^\mu  e_b{}^\nu\widehat{R}_{\mu\nu}{}^A\,,\qquad
 \widehat{R}_{\mu\nu}{}^A= \partial_\mu B_\nu{}^A - \partial_\nu B_\mu{}^A + B_\nu{}^C B_\mu{}^B f_{BC}{} ^A - 2B_{[\mu}^B\mathcal{M}_{\nu]B}{}^A\,.
 	\label{covcurvature}
 	\end{equation}
 \end{tcolorbox}
\medskip

A useful change of basis of transformations is to consider covariant general coordinate transformations (cgct), rather than gct. This is explained in detail in \cite[Sec. 11.3.2]{Freedman:2012zz}. They are defined as a combination of gct and other gauge transformations:
  \begin{equation}
\delta_{\rm cgct}(\xi)=\delta_{\rm gct}(\xi)-\delta(\xi^\mu B_\mu).
\label{defcgct}
\end{equation}
They act on the different fields as
\begin{align}
   \delta_{\rm cgct}(\xi)\phi =&\xi ^a{\cal D}_a\phi\,,\nonumber\\
\delta_{\rm cgct}(\xi)B_\mu{}^A=& \xi ^\nu \left( \widehat{R}_{\nu \mu }{}^A -B_{\mu }{}^B{\cal M}_{\nu
  B}{}^A\right) \,,\nonumber\\
   \delta_{\rm cgct}(\xi)e_\mu^a=&\partial _\mu \xi ^a  + \xi ^cB _\mu {}^Bf_{Bc}{}^a-\xi ^\nu R_{\mu\nu}{}^a\,,
 \label{cgctonfields}
\end{align}
where in the last equation an index $a$ on structure functions indicates the non-zero commutator between translations and other standard gauge transformations like Lorentz rotations ($[T_B,P_c] = f_{Bc}{}^a P_a$). The last term of (\ref{cgctonfields}) contains the curvature of translations, which often is put to zero as second-order formulation for the spin connection, which is defined in this way in function of other fields.

These cgct thus have only a derivative on the parameter for their gauge field $e_\mu {}^a$. Gct appear in commutators of standard gauge transformations in the form of cgct, and thus (also) don't have a derivative on the parameters when applied to covariant fields and gauge fields $B_\mu {}^A$.
\medskip

One further property is valid for elementary covariant fields (not necessary for a function thereof that is general covariant)
\begin{tcolorbox}[title=Transformation of covariant elementary field is covariant]
	If $\Phi^i$ is a covariant matter field and the algebra contains no derivatives on parameters:
	 $\delta(\epsilon)\Phi^i$ is a covariant quantity.
\end{tcolorbox}
To prove this property,
we use that the commutator of local symmetry transformations on these fields should not lead to derivatives on the parameter. The algebra is symbolically written as\footnote{In the commutator of standard gauge transformations (like two supersymmetries) cgct can appear. The remark after (\ref{cgctonfields}) that cgct only lead to derivatives on the parameter when applied to the frame field is thus relevant here. Derivatives on the parameters do not appear when cgct are applied to fields $\Phi ^i$ and $B_\mu {}^A$.}
\begin{equation}
  \left[\delta (\epsilon _1),\delta (\epsilon _2)\right]\Phi ^i = \delta (\epsilon _3)\Phi ^i + \epsilon _2^B\epsilon _1^A\eta _{BA}^i\,,\qquad \epsilon _3^A= \epsilon _2^B\epsilon _1^C f_{CB}{}^A\,,
 \label{algebraonphi}
\end{equation}
where $\eta _{BA}^i$ are the non-closure functions (which should be proportional to field equations) and $f_{AB}{}^C$ are the structure functions. \emph{We assume that there is no spacetime derivative on the parameters $\epsilon _1$ and $\epsilon _2$}, as is usually the case.

We thus start from the transformation of $\Phi ^i$ under $\epsilon _2$, and since $\Phi ^i$ is covariant:
\begin{equation}
  \delta (\epsilon _2)\Phi ^i = \epsilon _2^A R_A{}^i(\phi ,\partial \phi,\ldots  )\,,
 \label{delphi2}
\end{equation}
 does not contain a derivative on the parameter $\epsilon _2$. Hence in $\delta (\epsilon _1)\delta (\epsilon _2)\Phi ^i$ there are no spacetime derivatives on $\epsilon _2$. Since there are also no such derivatives in what should come out of the commutator, i.e. (\ref{algebraonphi}), there should not be derivatives on $\epsilon_2$ when calculating $\delta (\epsilon _2)\delta (\epsilon _1)\Phi ^i$. Thus $R_A{}^i$ should be a covariant quantity.

We now consider this requirement explicitly. Following (\ref{DefDAmucompact}) this is
\begin{equation}
D^\nu _B R_A{}^i=R_B{}^j H_{jA}{}^{\nu i}+\frac{\pl R_A{}^i}{\partial B_\nu {}^B}=0\,.
 \label{covRArequirementH}
\end{equation}
When we take the derivative w.r.t. $\partial _\mu \phi ^k$ of (\ref{covRArequirementH}), we get
\begin{equation}
  H_{kB}{}^{\mu j}H_{jA}{}^{\nu i} + (-)^{(B+j)k} R_B{}^j \frac{\pl H_{jA}{}^{\nu i}}{\partial \partial _\mu \phi ^k} + (-)^{Bk}\frac{\pl H_{kA}{}^{\mu i}}{\partial B_\nu {}^B}=0\,.
 \label{fullHHeqn}
\end{equation}
Since $H_{jA}{}^{\nu i}$ is defined as a derivative of $R_A^i$ w.r.t. $\partial _\nu \phi ^j$, the expression in the second term satisfies
\begin{equation}
  \frac{\pl H_{jA}{}^{\nu i}}{\partial \partial _\mu \phi ^k} = (-)^{jk}\frac{\pl H_{kA}{}^{\mu i}}{\partial \partial _\nu \phi ^j}\,.
 \label{symmetrydH}
\end{equation}
Therefore, we can write (\ref{fullHHeqn}) as
\begin{equation}
  H_{kB}{}^{\mu j}H_{jA}{}^{\nu i}  + (-)^{Bk}D^\nu _B H_{kA}{}^{\mu i}=0\,.
 \label{HHrewritten}
\end{equation}
This equation is important for the proof in the main part of the paper.

\section{Constructing the covariant eom for the chiral multiplet}
\label{app:exchiralg}
To illustrate the abstract quantities and the integrability condition (\ref{integrability}), we use the example of section \ref{ss:exchiral}.
We will consider the conformal gauge fields as background, and concentrate on the fields of the chiral multiplet. Thus, the fields are
\begin{equation}
  \{\phi ^i\} = \{ X, \, P_L\Omega ^\alpha , \,F\} \,,\qquad {i}={ \uX, \ualpha, \uF}\,.
 \label{phiichiral}
\end{equation}
We use again underlined indices to indicate values of $i$. The transformation that we consider is supersymmetry, and thus the index $A$ of the abstract notation is here $\alpha $.

Using the transformation laws (\ref{chiralconfgauged}) and the notation  (\ref{defHgeneralb}) we find the following non-zero coefficients $H_{jA}{}^{\mu i}$, which determine also the elements of $(L_\alpha ^\mu  \unity )_i{}^j$, see (\ref{Lunity}):
\begin{align}
  H_{\ubeta \alpha }{}^{\mu \uF}=&-(L_\alpha ^\mu  \unity )_{\ubeta}{}^{\uF}= -\frac{1}{\sqrt{2}}(P_L\gamma ^\mu )_{\beta \alpha }
  \,,\nonumber\\
  H_{\uX\alpha}{}^{\mu \ubeta }= &(L_\alpha ^\mu  \unity )_{\uX}{}^{\ubeta }   = -\frac{1}{\sqrt{2}}(P_R\gamma ^\mu )_\alpha {}^\beta \,,\nonumber\\
  H_{\uX\alpha}{}^{\mu \uF}=&(L_\alpha ^\mu  \unity )_{\uX}{}^{\uF}=-\frac12(P_R\gamma ^\nu \gamma ^\mu \psi _\nu )_\alpha \,.
 \label{Hchiral}
\end{align}
Only the last one, $H_{\uX\alpha}{}^{\mu \uF}$, depends on the gauge field and the only non-zero square of elements of $H$ can be built from $H_{\uX\alpha}{}^{\mu \ubeta }H_{\ubeta \alpha }{}^{\mu \uF}$.  Therefore the only non-vanishing element of $(L_\beta ^\nu \,L_\alpha ^\mu \unity )_i{}^j$ is according to the definition (\ref{defoperatorL})
\begin{align}
  (L_\beta ^\nu \,L^\mu  _\alpha \unity )_{\uX}{}^{\uF}&= \frac{\partial (L_\alpha ^\mu  \unity )_{\uX}{}^{\uF} }{\partial \psi _\nu ^\beta }-(L_\alpha ^\mu  \unity )_{\uX}{}^{\ugamma }H_{\ugamma \beta  }{}^{\nu \uF}\nonumber\\
  &=\ft12(P_R\gamma ^\nu \gamma ^\mu)_{\alpha \beta }-\ft12(P_R\gamma ^\mu \gamma ^\nu )_{\alpha\beta }= -(P_R\gamma ^{\mu\nu} )_{\alpha\beta }\,.
 \label{L2unity}
\end{align}
This is covariant,\footnote{In principle, for covariant objects we should go to flat indices $a,b,\ldots $, but this is a straightforward procedure.} and thus there are no non-vanishing third powers of $L$ on $\unity $, and $g_{\chi }=0$ for $\chi \geq 3$. We thus find from  (\ref{sol}) the non-vanishing components of $(g_{\chi })_i{}^j$
\begin{align}
  g_0 & =\unity \,, \nonumber\\
  (g_1)_{\ubeta}{}^{\uF} & =-\psi _\mu ^\alpha (L_\alpha ^\mu  \unity )_{\ubeta}{}^{\uF}= -\psi _\mu ^\alpha \frac{1}{\sqrt{2}}(P_L\gamma ^\mu )_{\beta \alpha } = \frac{1}{\sqrt{2}}(P_L\gamma ^\mu \psi _\mu )_\beta \,,\nonumber\\
  (g_1)_{\uX}{}^{\ubeta }& =-\psi _\mu ^\alpha(L_\alpha ^\mu  \unity )_{\uX}{}^{\ubeta }= \psi _\mu ^\alpha\frac{1}{\sqrt{2}}(P_R\gamma ^\mu )_\alpha {}^\beta =\frac{1}{\sqrt{2}} (\bar \psi _\mu P_R \gamma ^\mu )^\beta \,,\nonumber\\
  (g_1)_{\uX}{}^{\uF}&=-\psi _\mu ^\alpha (L_\alpha ^\mu  \unity )_{\uX}{}^{\uF}= \psi _\mu ^\alpha\frac12(P_R\gamma ^\nu \gamma ^\mu \psi _\nu )_\alpha=\frac12 \bar \psi _\mu P_R\gamma ^\nu \gamma ^\mu \psi _\nu \,,\nonumber\\
  (g_2)_{\uX}{}^{\uF}&= \frac12\psi  _\mu ^\alpha \psi _\nu ^\beta (L_\beta ^\nu \,L^\mu  _\alpha \unity )_{\uX}{}^{\uF}= \frac12\bar \psi  _\mu P_R\gamma ^{\mu\nu} \psi _\nu \,.
\label{gchicoeff}
\end{align}
Summing the last two gives
\begin{equation}
  g_\uX{}^\uF= \ft12\bar \psi _\mu P_R\psi ^\mu \,.
 \label{g2total}
\end{equation}

We thus obtain
\begin{align}
\Theta _\uF &= T_\uF\,,\nonumber\\
  \Theta _\ubeta &= T_\ubeta +\frac{1}{\sqrt{2}}P_L\gamma ^\mu \psi _\mu T_\uF\,,\nonumber\\ 
  \Theta _\uX&= T_\uX + \frac{1}{\sqrt{2}}\left(\bar \psi _\mu P_R \gamma ^\mu\right)^\beta  T_\ubeta +\frac12\bar \psi _\mu P_R\psi ^\mu T_\uF\nonumber\\
  &=T_\uX + \frac{1}{\sqrt{2}}\left(\bar \psi _\mu P_R\gamma ^\mu\right)^\beta \Theta _\ubeta -\frac12\bar \psi _\mu P_R\gamma ^{\mu \nu }\psi _\nu T_\uF\,,
 \label{Thetachiral}
\end{align}
which is the structure that we saw in  (\ref{XL1fe}).


\providecommand{\href}[2]{#2}\begingroup\raggedright\endgroup

\end{document}